# Whether the excited cluster $^{14}$N$^*$ exists in $^{15}$O nucleus?

S.B. Dubovichenko[a,b,1,*], N.A. Burkova[b,2], A.V. Dzhazairov-Kakhramanov[a,3,*]

[a] *Fesenkov Astrophysical Institute "NCSRT" ASA MDASI RK, 050020, Almaty, Kazakhstan*
[b] *al-Farabi Kazakh National University, MES RK, 050040, Almaty, Kazakhstan*

*E-mail addresses:* [1]dubovichenko@mail.ru, [2]natali.burkova@gmail.com, [3]albert-j@yandex.kz.

**Abstract**

Within the modified potential cluster model with forbidden states effectively taking into account the Pauli Exclusion Principle the astrophysical *S*-factor of radiative $p^{14}$N capture to $^{15}$O ground state has been calculated. The calculation was performed at the proton energy up to 5 MeV in the center of mass system (c.m.) with taking into account broad resonances up to 3.4 MeV in c.m. To explain reasonably the available experimental data it is required to admit the existence of $^{14}$N cluster in excited state $^{14}$N$^*$ with the excitation energy equal to 5.69 MeV and momentum $J^\pi = 1^-$. It is assumed that in this case the $^4D_{1/2}$ state wave function of the $p^{14}$N$^*$ clusters relative motion may be used. There was shown that it is succeeded to describe the *S*-factor of $p^{14}$N capture in the resonance region only under the assumption that all low-lying resonances at 260(1/2$^+$), 987(3/2$^+$), 1447(1/2$^+$), 2187(3/2$^+$), and 3211(3/2$^+$) keV in c.m. are the $^4D_{1/2}$ and $^4D_{3/2}$ scattering waves.

## 1. Introduction

Considering reaction of the proton radiative capture on $^{14}$N to the different bound states (BS) of $^{15}$O takes a part in CNO cycle of thermonuclear processes, which in many ways determined energy release processes of our Sun and, possibly, various stars of our Galaxy. Apparently, sufficient modern description of the astrophysical supplements, related with the proton radiative capture on $^{14}$N is given in review [1].

In particular, it was shown that globular clusters represent the oldest stellar populations. Their age coincides with the creation time elapsed since the epoch of formation of the first stars in the universe and provides an independent check of the reliability of standard (and non-standard) cosmological models. The main-sequence stars presently observed in globular clusters have masses smaller than that of the Sun. These low-mass stars burn in its center, mainly through the thermonuclear *pp* chain.

However, at the end of their life, when the central hydrogen mass fraction becomes smaller than about 0.1, the nuclear energy released by the *H*-burning becomes insufficient and the stellar core contracts to extract energy from its own gravitational field. Then, the central temperature and density increase and the star burning switches from the *pp* chain to the more effective CNO cycle. However, exactly considered here reaction $^{14}$N(p,γ)$^{15}$O is the bottleneck of the CNO cycle. There was long standing problem between observation and stellar models. We can see over-production of all chemical elements above carbon by a factor of 2 or more. This mismatch has been resolved by using the lower reaction rate of $^{14}$N($p,\gamma$)$^{15}$O for processes of evolution of chemical elements [1].

Thus, it is reputed that the considered here reaction $^{14}$N($p,\gamma$)$^{15}$O has a big sense for very different problems and studying its mechanism plays a big role for nuclear physics, because meanwhile we do not succeed to describe the capture to the ground state (GS) in the framework of some model. This reaction, as was shown above, plays the big role for nuclear astrophysics too and its consideration in the frame of nuclear physics methods should to find out the real shape and rate value of this reaction.

To continue studying of the radiative capture process [2,3] $p^{14}$N → $^{15}$O$\gamma$ with the capture to the GS of $^{15}$O presented in the $p^{14}$N channel has been considered within the framework of the modified potential cluster model (MPCM) with forbidden states (FS) which effectively take into account the Pauli Exclusion Principle [4,5]. This model has already been used to study thirty radiative capture reactions on light nuclei at astrophysical and thermal energies [2,3,5]. The given model enables to describe reasonably the basic characteristics of these reactions if intercluster potentials contain the bound forbidden states. As a rule, the model has four parameters the values of which are fixed on the basis of description of the characteristics of scattering and the bound states before the consideration of the radiative capture processes.

---
[*] Corresponding authors

Before the considering the obtained results it should be noted that the only attempt to explain qualitatively the failures in description of the proton capture on $^{14}$N to the GS of $^{15}$O has been made in [6] on the basis of the shell model (SM). Evidently, it is the only work in which the ground state "problem" is discussed within the SM structural peculiarities. Notably, in the intermediate coupling model the GS wave function is presented in the form of superposition of two components $A \cdot [^{14}N(1_1^+) \otimes p_{3/2}] + B \cdot [^{14}N(1_1^+) \otimes p_{1/2}]$ amplitudes of which may be varied. However, in this work it is not succeeded describing the experimental $S$-factor for the capture to the GS of $^{15}$O either, though all known data for the capture to the excited states (ES) are described reasonably enough.

Moreover, [7] presents the calculation results of more than thirty processes such as the protons radiative capture and other reactions at astrophysical energies in the potential model (PM) including $p^{14}N \to {}^{15}O\gamma$ process with capture to the GS and all subthreshold levels of $^{15}$O. It seems reasonable to say that such calculation is substantially the fitting of experimental data because each of partial cross-sections is normalized on its resonance value, and then the total cross-section is considered as being reliable. There are two more works in one of which the capture to the fourth ES of $^{15}$O was considered within the potential model [8], and [9], where the same capture was studied, but the interparticle potentials of 16 reactions considered were built on the basis of the M3Y potentials folding model [10].

Nevertheless we failed to find any theoretical calculations of the astrophysical characteristics of this reaction with the capture to the GS within the resonating groups or "*ab initio*" methods. It seems reasonable to say that practically all available works except ones given above are restricted to only the $R$-matrix calculation of available experimental data for the captures to the ground and excited states of $^{15}$O. Note that the $R$-matrix analysis results obtained in various works, in spite of tens of the fitting parameters, differ greatly from each other due to the experimental data ambiguity (see for example [11,12] and references given in these works to prior publications).

## 2. Calculation methods

In the present calculations we used the cluster states classification according to Young tableaux [4] as it was done for all reactions considered earlier [2.3.5]. In this case we suppose that for $^{14}$N the orbital Young tableau may be presented in the form $\{f\}_L = \{4442\}$, then for the $p^{14}$N system within the framework of 1$p$-shell we have $\{f\}_L = \{1\} \times \{4442\} \to \{5442\} + \{4443\}$ [13]. The first of the obtained tableaux is compatible with the orbital momenta $L = 0$ and 2, and it is forbidden because there cannot be five nucleons in the $s$-shell [4]. The second one is allowed and compatible with the orbital momentum $L = 1$. Therefore, there are the forbidden bound states for the $p^{14}$N system in the potentials of the $S$ and $D$ scattering waves. At the same time the $P_{1/2}$ wave has only the allowed state (AS) that may correspond to $^{15}$O GS with $J^{\pi} = 1/2^-$ and have the $p^{14}$N system binding energy equal to -7.2971 MeV if the $^{14}$N cluster is in the ground state [14].

All basic expressions for calculations used within the framework of the MPCM are given in [2,3,5,15]. For the nuclear potentials depending explicitly on momenta $JLS$ and the Young tableaux $\{f\}$ we usually use the Gauss type form

$$V_{JLS\{f\}}(r) = -V_{0,JLS\{f\}} \exp[-\alpha_{JLS\{f\}} r^2] + V_{\text{coul}}(r) \qquad (1)$$

with the point-like Coulomb term [2,3,15] given below. In the current calculations there were used the following values of particles masses $m_p = 1.007276469$ amu [16] and $m(^{14}N) = 14.003074$ amu [17], the constant $\hbar^2/m_0$ ($m_0$ is atomic mass unit) value is equal to 41.4686 MeV·fm². The Coulomb parameter was defined as $\eta = 3.44476 \cdot 10^{-2} \cdot Z_1 Z_2 \mu/q$, where $q$ is the wave number in fm⁻¹ determined by the energy of interacting particles in the initial channel. The Coulomb potential for zero Coulomb radius $r_{Coul} = 0$ was written as $V_{Coul}$(МэВ) = $1.439975 \cdot Z_1 Z_2/r$, where $r$ in fm is the relative distance between the initial channel particles. For the clusters magnetic momenta there were used the values $\mu_p = 2.792847\mu_0$ [16] and $\mu(^{14}N) = 0.404\mu_0$ [17], where $\mu_0$ is the nuclear magneton.

## 3. Results

First of all let us present the spectrum of resonance states at positive energies, which have some influence on the processes of the radiative proton capture on $^{14}$N, have comparatively large widths, and will be taken into account in the present calculations. These results were taken from Tables 15.16 and 15.20 of [14] although there are already minor refinements of some of these data, presented in [18]:

1. The first resonance state of $^{15}$O in the $p^{14}$N channel is at the energy equal to 259.4(4) keV. Note that all energies in this letter are given in c.m. This state has the width 0.99(10) keV and the momentum $J^\pi = 1/2^+$. Such resonance may be conformed to the $^2S_{1/2}$ scattering wave with the FS [3], though, of course, this state may conform also to the quartet $^4D_{1/2}$ resonance wave with the FS.

2. The second resonance state has the energy equal to 986.9(5) keV, its widths is equal to 3.6(7) keV and the momentum $J^\pi = 3/2^+$. Such state may be conformed to the $^4S_{3/2}$ scattering wave with the FS, however it may be also the mixed doublet $^2D_{3/2}$ and quartet $^2D_{3/2}$ resonance scattering waves.

3. The third resonance state has the energy equal to 1447(6) keV, its width is equal to 32 keV and the momentum $J^\pi = 1/2^+$. It may comply with the $^2S_{1/2}$ scattering wave with the FS, however this state may be also the quartet $^4D_{1/2}$ resonance scattering wave.

4. The fourth resonance state has the energy equal to 2187(8) keV, its width is equal to 191 keV and the momentum $J^\pi = 3/2^+$. It may be conformed to the $^4S_{3/2}$ scattering wave with the FS or to the mixed doublet $^2D_{3/2}$ and quartet $^4D_{3/2}$ resonance scattering waves.

5. The fifth resonance state has the energy equal to 3211 keV, its width is equal to 140(40) keV and the momentum $J^\pi = 3/2^+$. It may be conformed to the $^4S_{3/2}$ scattering wave with the FS or to the mixed doublet $^2D_{3/2}$ and quartet $^4D_{3/2}$ resonance scattering waves.

We did not consider the resonances at higher energies because the experimental data were obtained only up to 3.4 MeV [19], and the next broad resonance is at 3.6 MeV [14]. Here we suppose that all scattering resonances described above are in the $D$ waves, so for the nonresonance $^2S_{1/2}$ and $^4S_{3/2}$ scattering waves there were obtained the potentials parameters leading to near-zero scattering phase shifts. Since these waves contain the FS, the potential depth has nonzero value [2,3,15]. These parameters values do not play the essential role, only the near-zero values of their scattering phase shifts are important.

Moreover, at the energy up to 2.5–3.0 MeV above the threshold in $^{15}$O spectra there are not any resonance levels with negative parity and large width, which might correspond to resonances in the $P$ waves of scattering [14]. Thus, the phase shifts of these partial waves may be taken as equal or approximate to zero, and since there are not any bound FS in the $P$ waves [2,3,15], the depth of these potentials may be simply taken as equal to zero. The parameters of all scattering potentials used further are given in Table 1. The potentials of the resonance waves were built to describe correctly the resonance energy and its width.

Note that we did not manage to describe correctly the astrophysical $S$-factor of the proton capture on $^{14}$N to the GS of $^{15}$O considering this nucleus in the two-particle $p^{14}$N channel and taking into account the $^{14}$N cluster being in the GS. Any parameters of the GS potential in the $^{2+4}P_{1/2}$ wave for the $p^{14}$N particles relative motion in $^{15}$O do not enable to describe correctly available experimental data. The MPCM that enabled previously to describe correctly the total cross-sections of more than 30 radiative capture reactions [2,3,5,15], in this case led to the results similar to [6]. For example, at the resonance energies equal to 260 and 987 keV the astrophysical *S*-factor was 4 orders more than available experimental data, and the results for the *S*-factor in the first resonance at 260 keV did not depend practically on what kind of wave $^2S_{1/2}$ or $^4D_{1/2}$ was used for it.

**Table 1.**
The $p^{14}$N scattering potentials parameters. The third and fourth columns present the parameters for the Gauss potential (1), the fifth and sixth ones shows the resonance energy and its width correspondingly.

| No. | Scattering states | $V_0$, MeV | $\alpha$, fm$^{-2}$ | $E_{cm}$, keV | $\Gamma_{cm}$, keV |
|---|---|---|---|---|---|
| 1 | 1st resonance in $^4D_{1/2}$ | 20.029545 | 0.013 | 260(1) | 0.97(1) |
| 2 | 2nd resonance in $^4D_{1/2}$ | 571.622 | 0.25 | 1447(6) | 33.6(1) |
| 3 | 1st resonance in $^4D_{3/2}$ | 915.61367 | 0.4 | 987(1) | 3.1(1) |
| 4 | 2nd resonance in $^4D_{3/2}$ | 474.42 | 0.21 | 2187(1) | 197(1) |
| 5 | 3rd resonance in $^4D_{3/2}$ | 1463.247 | 0.65 | 3211(1) | 131(1) |
| 6 | Nonresonance $^2S_{1/2}$ and $^4S_{3/2}$ | 3500.0 | 10.0 | --- | --- |

At the same time the assumption that $^{14}$N in $^{15}$O is in the excited state $^{14}$N* with the excitation energy equal to 5.69 MeV and momentum $J^\pi = 1^-$, enabled to reproduce reasonably the available experimental data on the proton capture on $^{14}$N to the GS of $^{15}$O. Admit that after proton capture on $^{14}$N and emission of $\gamma$ quantum with the energy equal to 7.2971 MeV, the GS of $^{15}$O transits immediately to the stationary $^4D_{1/2}$ state in the $p^{14}$N* channel. A part of the $p^{14}$N* system binding energy is expended on the $^{14}$N cluster excitation so that relative energy of the new $p^{14}$N* system becomes equal to -1.6057 MeV. Furthermore, we make an assumption that for the astrophysical *S*-factor calculations there may be used the wave function of that $^4D_{1/2}$ stationary state.

The parameters of this GS potential are presented in Table 2. Three last columns show the binding energy $E_b$, the dimensionless asymptotic constant (AC) $C_w$ determined usually as $\chi_L(r) = \sqrt{2k_0} C_w W_{-\eta L+1/2}(2k_0 r)$ [20], and the charge radius $<r^2>^{1/2}$ correspondingly. Since we failed to find information about the asymptotic constant in such GS, the potential parameters were fitted to describe reasonably the *S*-factor value at zero energy totally determined by the *E*1 transitions from the *P* scattering waves with zero potentials.

The *M*1 capture from the $^4D_{1/2}$ scattering wave to the $^4D_{1/2}$ bound state is turned out to be possible because for these states there were used different potentials. Each of such potentials describes its own set of characteristics inherent to only these states, and it is impossible to obtain for them the potential with the unified parameters. It may be caused, for example, by the difference between the Young tableaux for such states as it was considered earlier in [4,5]. Such tableau is considered as an additional quantum number, and if it has the different form for the continuous and discrete spectra, the potentials of these states may be different [4]. In this case the situation may be analogous because the given above consideration of the states according to Young tableaux has the qualitative character. It is caused by that we have not any tables for exact products of these tableaux in form [13].

**Table 2.**
The parameters of the potential for the ground $^4D_{1/2}$ state of $^{15}$O in the $p^{14}$N* channel.

| $V_0$, MeV | $\alpha$, fm$^{-2}$ | $E_b$, MeV | $C_w$ | $<r^2>^{1/2}$, fm |
|---|---|---|---|---|
| 134.246725 | 0.09 | -1.60570 | 3.0(1) | 2.9 |

**Table 3.**
Transitions to the $^4D_{1/2}$ ground state of $^{15}O$ in the $p^{14}N^*$ channel for $p^{14}N$ capture.

| № | $[^{(2S+1)}L_J]_i$ | Type of transition | $[^{(2S+1)}L_J]_f$ |
|---|---|---|---|
| 1 | $^4P_{1/2}$ | E1 | $^4D_{1/2}$ |
| 2 | $^4P_{3/2}$ | E1 | $^4D_{1/2}$ |
| 3 | $^4S_{3/2}$ | E2 | $^4D_{1/2}$ |
| 4 | $^4D_{3/2}$ | E2 | $^4D_{1/2}$ |
| 5 | $^4D_{1/2}$ | M1 | $^4D_{1/2}$ |
| 6 | $^4D_{3/2}$ | M1 | $^4D_{1/2}$ |

In the case of consideration of the proton capture on $^{14}N$ there were taken into account the radiative transitions between the initial $[^{(2S+1)}L_J]_i$ and final $[^{(2S+1)}L_J]_f$ states presented in Table 3. The results of calculations are shown in Fig. 1. The blue dashed line presents the capture processes No. 4, 5, 6 from Table 3. Capture No.3 is shown at the foot of Fig. 1 by the black dotted line. Its contribution to the total S-factor is practically small. The experimental data for the S-factor were taken from [18,19,21–24]. If the first resonance is admitted to exist in the $^2S_{1/2}$ scattering wave, there would not be the first resonance in the S-factor because in the cluster model the transitions between the states being in the different spin channels are forbidden [2,3].

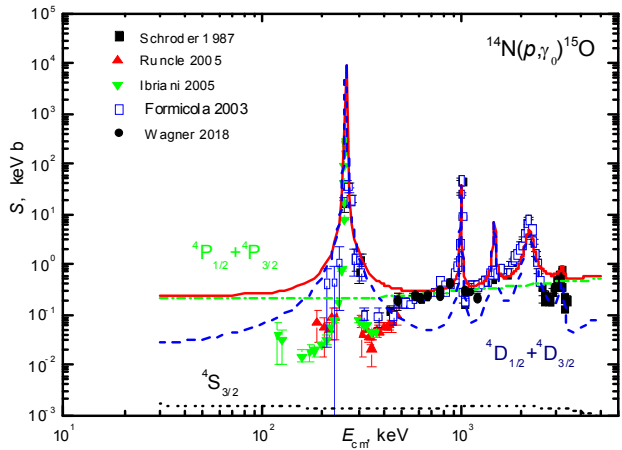

**Fig.1.** The astrophysical S-factor of the radiative $^{14}N(p,\gamma)^{15}O$ capture to the GS of $^{15}O$ in the $p^{14}N^*$ channel. Points correspond to the experimental data: black squares [19], downward green triangles [21], upward red triangles [22], open blue squares [18,23], and grey circles [24]. The green dotted-dashed line shows the calculation results for the E1 transitions from the $^4P_{1/2}$ and $^4P_{3/2}$ waves. The blue dashed line corresponds to the calculation of the E2 and M1 transitions from the $^4D_{1/2}$ and $^4D_{3/2}$ waves. The black dotted line shows the calculation of the E2 transition from the $^4S_{3/2}$ wave. The red solid line presents the total S-factor.

At the energy equal to 30 keV the S-factor with such potential of the GS is equal to 0.24 keV b, and at the energy range from 30 keV to 110 keV its values are in the range of 0.26(2) keV b. The calculated value of the first resonance, as it can be seen from Fig. 1, is slightly overestimated in comparison with the experimental data, i.e., it is equal to $9.5 \cdot 10^3$ keV·b against the experimental one equal to 292 keV b [21]. The width of this resonance is reproduced correctly but in the range of the smallest energies the calculated S-factor is much greater than data [21,22], though its value at zero energy is in a good agreement with other results given in Table 4. From the results of this table for the S-factor we obtain the range of values from 0.08 to 0.57 keV b, that gives the value averaged over the range, equal to 0.33(25) keV b. It follows that the spread of values is too large to draw definite conclusions about the S-factor value at zero energy.

**Table 4.**
The S-factor values obtained in the different works and their average value.

| References | [25] | [21] | [22] | [26] | [27] | [28] | [24] | Average |
|---|---|---|---|---|---|---|---|---|
| S(0), keV b | 0.25(5) | 0.25(6) | 0.49(8) | 0.27(5) | 0.15(7) | 0.42(4) | 0.19(5) | 0.29 |

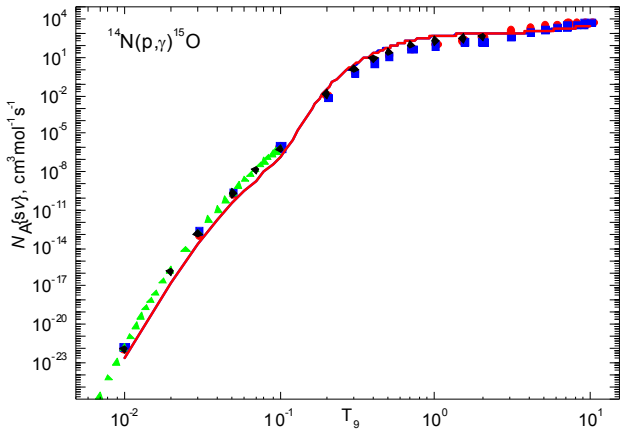

**Fig. 2.** The rate of the radiative $^{14}N(p,\gamma)^{15}O$ capture. Points show the results for the total reaction rate with the capture to all bound states of $^{15}O$: black rhombus [11], red circles [25], upward green triangles [27], blue squares [29]. Our results for the capture to the GS are shown by the red solid line. The approximation of our results by the analytical form (3) is absolutely coincides with calculation results shown by the red solid.

Furthermore, Fig. 1 illustrates that we managed to describe well the second maximum equal to 40 keV b at 990 keV against its experimental value equal to 48 keV b [23]. In the range of the energies between the resonances the calculated *S*-factor is in good agreement with data [23], especially in the region of the second resonance, although it has not the sharp decreasing observed in the experiment after the first resonance. The *S*-factor value between the resonances is totally determined by the *E*1 transitions from the *P* scattering waves with zero potentials.

The third resonance is one order of magnitude greater than the available data, but the fourth one is in good agreement with results [23]. It can be seen from Fig. 1 that there are not any experimental data for the third resonance in the $^4D_{1/2}$ wave at 1447 keV. Although there is observed some increasing of the *S*-factor in the region of this resonance, nevertheless, there were not performed any measurements of the cross-section exactly at the resonance energy [19,23]. The fifth resonance is also observed, and its height is in good agreement with data [19], but the description of the minimum between this resonance and the previous one is comparatively worse. Such result is also caused by the *S*-factor behavior at the *E*1 capture from the *P* scattering waves (see the green dotted-dashed line in Fig. 1). The potential of these waves was assumed to be equal to zero, but at 2-3 MeV this assumption is no longer correct. In order to build correctly the potential we need the phase shift analysis results for the elastic $p^{14}$N scattering.

**Table 5.** The parameters for analytical parameterization (3) of the reaction rate.

| No. | $a_i$ |
|---|---|
| 1 | -0.2947718E+03 |
| 2 | 0.1475694E+02 |
| 3 | -0.6526511E+04 |
| 4 | -0.6565516E+04 |
| 5 | -0.1623549E+04 |
| 6 | -0.4949766E+02 |
| 7 | 0.1806048E+03 |
| 8 | 0.1312910E+03 |
| 9 | 0.4685491E+06 |
| 10 | 0.1547907E+02 |
| 11 | 0.5074225E+04 |
| 12 | 0.2791499E+01 |
| 13 | 0.1733075E+03 |
| 14 | 0.1121820E+01 |
| 15 | 0.5450381E+03 |
| 16 | 0.1532875E+01 |

If the reaction total cross-sections or the radiative capture process astrophysical *S*-factor are known, there may be determined this reaction rate [29]

$$N_A \langle \sigma v \rangle = 3.7313 \cdot 10^4 \mu^{-1/2} T_9^{-3/2} \int_0^\infty \sigma(E) E \exp(-11.605 E/T_9) dE, \quad (2)$$

where *E* is the energy measured in MeV, the cross-section $\sigma(E)$ is measured in µb, µ is the reduced mass in amu, and $T_9$ is the temperature in units of $10^9$ K. The calculation results for the reaction rate in the case of capture to the GS are shown in Fig. 2 by the blue solid line. For this calculation there were used the shown in Fig. 1 *S*-factor values from 10 keV to 5 MeV with a step value of 10 keV. Since the *S*-factor was calculated from 30 keV, in the first two points at 10 and 20 keV its value at 30 keV was used. Such assumption is well admitted because the *S*-factor values below 100 keV are practically constant. Fig. 2 illustrates that in the temperature range 0.2–2.0 $T_9$ our results for the rate of the capture to the GS is slightly overestimated in comparison with other works, where the total rate of the proton capture on $^{14}$N reaction to all BS of $^{15}$O is presented.

Further there was performed the parameterization of the calculated reaction rate by the form [30]

$$N_A \langle sv \rangle = \frac{a_1}{T^{2/3}} \exp(-\frac{a_2}{T^{1/3}})(1 + a_3 T^{1/3} + a_4 T^{2/3} + a_5 T + a_6 T^{4/3} + a_7 T^{5/3} + a_8 T^{7/3}) +$$
$$+ \frac{a_9}{T^{1/2}} \exp(-\frac{a_{10}}{T^{1/2}}) + \frac{a_{11}}{T} \exp(-\frac{a_{12}}{T}) + \frac{a_{13}}{T^{3/2}} \exp(-\frac{a_{14}}{T^{3/2}}) + \frac{a_{15}}{T^2} \exp(-\frac{a_{16}}{T^2}). \quad (3)$$

The expansion parameters values are presented in Table 5, the quantity $T_9$ is denoted by *T*, and the calculated reaction rate error for $\chi^2$ calculation was taken equal to 5%. The reaction rate parameters from Table 5 lead to the average value of $\chi^2$ equal to 0.13, and the calculation results obtained by using this formula absolutely coincides with calculation results shown in Fig. 2 by the red solid line.

## 4. Conclusion

Thus, the assumption about the $^{14}$N* cluster excited state existing at the energy equal to 5.69

MeV with the momentum $J^{\pi} = 1^-$ in $^{15}$O and using for it the ground $^4D_{1/2}$ state enables in whole to describe correctly the experimental data for the astrophysical *S*-factor of the radiative $p^{14}$N capture to the GS in the range of low energies. However, as it was noted above, the experimental data errors are very big especially in the range of low energies. Therefore, it may be hoped that new measurements of the cross-sections will help to avoid the existing ambiguities including the one in the range of the third resonance at 1447 keV [31].

**Acknowledgments**

This work was supported by the Ministry of Education and Science of the Republic of Kazakhstan (Grant No. BR05236322 entitled "Study reactions of thermonuclear processes in extragalactic and galactic objects and their subsystems"), through the Fesenkov Astrophysical Institute of the National Center for Space Research and Technology of the Ministry of Defense and Aerospace Industry of the Republic of Kazakhstan (RK).